\documentclass[useAMS, usenatbib]{mn2e} 

\usepackage{graphicx,amssymb}

\title{A long-wavelength instability involving the stress tensor}

\author[K.M.~Schure \& A.R.~Bell]{K.~M.~Schure$^{1}$\thanks{E-mail: k.schure1@physics.ox.ac.uk}
, A.~R.~Bell$^{1}$\\
$^{1}$Department of Physics, University of Oxford, Clarendon Laboratory, Parks Road, Oxford OX1 3PU, United Kingdom}

\begin{document}

\newcommand\araa{{ARA\&A}}
\newcommand\apj{{ApJ}}
\newcommand{\apjl}{ApJL}
\newcommand\apjs{{ApJS}}
\newcommand\aap{{A\&A}}
\newcommand\mnras{{MNRAS}}
\newcommand\rmxaa{{Rev. Mexicana Astron. Astrofis.}}
\newcommand\nat{{Nature}}
\newcommand\physrep{{Phys.~Rep.}}
\newcommand\ssr{{Space Sci. Rev.}}
\newcommand\vv{{\bf v}}
\newcommand\fv{{\bf f}}
\newcommand\uv{{\bf u}}
\newcommand\Ev{{\bf E}}
\newcommand\Bv{{\bf B}}

\date{\ldots; \ldots}
\pagerange{\pageref{firstpage}--\pageref{lastpage}}\pubyear{2011}
\maketitle
\label{firstpage}

\begin{abstract}
Cosmic ray acceleration through first-order Fermi acceleration in a collisionless plasma relies on efficient scattering off magnetic field fluctuations. Scattering is most efficient for magnetic field fluctuations with wavelengths on the order of the gyroradius of the particles. In order to determine the highest energy to which cosmic rays can be accelerated, it is important to understand the growth of the magnetic field on these large scales. We derive the growth rate of the long-wavelength fluctuations in the linear regime, using the kinetic equation coupled to Maxwell's equations for the background plasma. The instability, driven by the cosmic ray current, acts on large scales due to the stress tensor and efficient scattering on small scales, and operates for both left- and right circular polarisations. This long-wavelength instability is potentially important in determining the acceleration efficiency and maximum energy of cosmic rays around shock waves such as in supernova remnants.
\end{abstract}

\begin{keywords}
instabilities ---  cosmic rays --- acceleration of particles --- ISM: supernova remnants
\end{keywords}

\section{Introduction}
\label{sec:intro}

The acceleration of Galactic cosmic rays up to energies of $10^{15}$~eV is widely believed to result from first order Fermi acceleration \citep{1977Axfordetal, 1977Krymskii, 1978Bella, 1978Bellb, 1978BlandfordOstriker} in supernova remnant blast waves. This mechanism relies on a converging flow, where cosmic rays gain energy every time they cross the discontinuity, and is also referred to as diffusive shock acceleration (DSA). On either side of the shock the cosmic rays isotropise by scattering off magnetic fluctuations, while the plasma is highly collisionless. The length scale of the magnetic fluctuations and the amplitude determine the scattering efficiency. The more quickly the particles isotropise, the quicker they may diffuse back accross the shock and the shorter the acceleration time.

There exists a variety of cosmic ray streaming instabilities that amplify the local magnetic field and increase the scattering efficiency of the plasma. For a recent review, see e.g.~\citet{2011Bykovetalreview}. 
On small scales, i.e.~smaller than the gyroradius, the streaming cosmic rays induce a return current that creates a force perpendicular to the background magnetic field. This results in a non-resonant instability that can amplify the magnetic field to many times its initial level in the nonlinear regime \citep{2004Bell}. Since scattering is most efficient when the larmor radius is of the order of the wavelength of the magnetic fluctuations, it is therefore expected that this instability is effective in confining the low-energy cosmic rays. 

Resonant instabilities arise in the regime where the parallel component of the wavelength corresponds to the gyroradius of the streaming cosmic rays \citep{1967Lerche,1969KulsrudPearce,1974Wentzel,1975Skillingc}. When the perturbed field becomes of the same order as the background magnetic field ($\delta B/B > 1$) the resonance is lost. The saturation level of the waves therefore is rather low, and this instability is not effective in amplifying the magnetic field to many times its zeroth order level. 

Whether the highest-energy cosmic rays are efficiently confined in the shock region depends on whether instabilities develop on scales larger than the gyroradius of the generating cosmic rays. 
In this paper we show that an instability acts in this regime that arises from the cosmic ray current and is mediated by the stress tensor. The coupling of the cosmic rays to the plasma on the small scales, due to effective scattering as a result of the non-resonant instability \citep{2004Bell}, further aids in making both left and right-handed polarisations unstable. 

We use the Boltzmann transport equation together with Maxwell's equations for the background plasma, to derive the growth rate of magnetic fluctuations on the large scale. Rather than limiting ourselves to the diffusion approximation, we include higher order anisotropy and find that the stress tensor takes an active part in mediating the instability. The small-scale scattering couples the two polarised modes and integrates the short-scale with the large-scale instability. The growth rate is found to steeply depend on the wavenumber, resulting in an rapid decline with wavelength.

We limit ourselves to a mono-energetic beam of particles. 
This will allow us to get a cleaner picture of how exactly the instability works. The distribution function generally is expected to follow a power law beyond the thermal peak, and has a cut-off at the high-energy end. If diffusion is a function of momentum, which is generally assumed, the high-energy cosmic rays diffuse farther upstream of the shock than the lower energy ones. The low-energy cut-off therefore increases with distance upstream of the shock. Since the number of particles steeply declines with energy, it can be argued that the instability is dominated by the cosmic rays with the lowest energy in the local plasma, being a function of distance ahead of the shock. 

In Section~\ref{sec:method} we show how we derive the dispersion relation for the instability that results from including the stress tensor and scattering on small scales. In Section~\ref{sec:k0limit} we evaluate the result and look at the growth rate of the instability for a range of collision efficiencies and wave numbers. We show the differences between inclusion or neglect of both the stress tensor and the small-scale diffusivity in Section~\ref{sec:diff_f1f2}. We discuss consequences for confinement of cosmic rays in Section~\ref{sec:resonance}. In Section~\ref{sec:f3} we show that in most regimes it suffices to limit ourselves to the second order anisotropy (i.e.~the stress tensor) and illustrate the difference when higher order anisotropy of the distribution function is taken into account. 
Finally, we discuss its implications and conclude in Section~\ref{sec:discussion}.

\section{Method}
\label{sec:method}
The distribution of relativistic particles can be described by a modified Boltzmann equation, also known as the Vlasov-Fokker-Planck equation, which reads (in cgs):
\begin{eqnarray}
\label{eq:kinetic}
\partial_t \fv + \vv \cdot \nabla \fv + \frac{q}{m}\left(\Ev + \frac{\vv \times \Bv}{c}\right)\cdot\nabla_v\fv = 
\nabla_v \cdot ({\bf D} \cdot \nabla_v \fv),
\end{eqnarray}
with 
$\fv$ the particle distribution in phase space, $\vv$ the particle velocity, and ${\bf D}= \frac{v^2\nu}{2}\left({\bf I} - {\bf \hat n}{\bf \hat n}\right)$ the diffusion tensor with $\nu$ the collision frequency. Here ${\bf I}$ is the identity matrix, and ${\bf \hat n}$ the unit vector in the direction of the corresponding tensor component. The effective collision frequency is a result of small angle scattering by fluctuations in the magnetic field.

The distribution function can be decomposed into an isotropic part and arbitrarily many moments, which sample the anisotropy to increasing order, as follows:
\begin{eqnarray}
\fv&=&f_0+\frac{f_i v_i}{v} + \frac{f_{ij} v_i v_j}{v^2} + \frac{f_{ijk} v_i v_j v_k}{v^3} +\ldots,\\\nonumber
v&=&\sqrt{\vv_x^2+\vv_y^2+\vv_z^2}
\end{eqnarray}
In spherical coordinates the directional components are given by:
\begin{eqnarray}
v_x & =&v \cos \theta\\\nonumber
v_y & =&v \sin \theta \sin \phi\\\nonumber
v_z & = &v \sin \theta \cos \phi
\end{eqnarray}

By integrating the distribution function over angle $\theta=[0,\pi], \phi=[0,2\pi]$ after multiplying with $\sin \theta$, the scalar transport equation can be derived (where we use $v=c$ for cosmic rays):
\begin{eqnarray}
\label{eq:f0}
4\pi\left[\partial_t f_0 + \frac{c}{3} \nabla \cdot \fv_1\right]  = 0.
\end{eqnarray}
The first moment of the distribution function can be found by mulitplying it with $\cos\theta, \sin\theta \sin\phi, \sin\theta \cos\phi$ to get the moment in $x,y,z$ respectively. 
The equation for momentum transport is thus found to be:
\begin{eqnarray}
\label{eq:f1}
\frac{4\pi}{3}\left[\partial_t \fv_1 + c\nabla\fv_0 + \frac{q}{mc}(\Bv \times \fv_1) +\nu \fv_1 + \frac{2}{5}c \nabla \cdot \{\fv_{2}\} = 0\right].
\end{eqnarray}

We restrict ourselves to a parallel shock and choose the direction of shock propagation to be in ${\bf \hat x}$, meaning that the background magnetic field only has a $B_x$ component (${\rm B_0}=|B_0|{\bf \hat x}$). We assume the electric field can be neglected.

The unperturbed cosmic ray current follows directly from the momentum transport equation when looking at the background quantities only, where we use that ${j}_{cr} = n_{cr} q u_s = n_{cr} q f_{x(0)} c/3$. The subscript 0 or 1 between brackets is used to indicate zeroth order and perturbed variables in case ambiguity can arise between orders in the expansion of the distribution function (no brackets) and orders in the perturbation (with brackets). A single index ($x,y,z$) is used for $\fv_1$ when a specific directional component is meant. 
Equation~\ref{eq:f1} gives for the zeroth order current in $\bf \hat x$:
\begin{eqnarray}
f_{x(0)}  &=& \frac{c}{\nu L_x} f_{0},
\end{eqnarray}
with $L_x$ the scale height of the cosmic ray density gradient that is determined by the diffusion coefficient over the shock velocity $D/u_s$. Using that $\nu=c^2/(3D)$ the zeroth order current is given by $f_{x(0)}=(3 u_s/c)f_0$. 

By eliminating $\fv_1$ between the first two moment equations, while neglecting higher order terms, the diffusion approximation is derived. Often the second- and higher order anisotropic parts are neglected, since each subsequent order in anisotropy is another factor $u_s/c$ smaller, where $u_s$ is the shock velocity. In this section however, we relax that assumption and include the next order in anisotropy ($\fv_2$). To higher order the transport equations can be derived by taking subsequent moments of the distribution function.

In systems with strong shock waves, such as for example SNR blast waves, it has been shown that the first order anisotropy, being effectively the cosmic ray current, is responsible for driving an instability on small scales \citep{2004Bell}. The cosmic ray current triggers a return current in the plasma. The ${\bf j}_{cr}\times \Bv_1$ force acts to focus the cosmic rays, where subscript 1 indicates perturbed variables. The resulting reaction force pushes out the plasma and magnetic field for right-hand polarised waves, resulting in a plasma instability that can act to amplify the magnetic field to many times its background level. Here we evaluate how the cosmic ray streaming instability develops when collisions and $\fv_2$ are explicitly taken into account.

The first order perturbation of Equation~\ref{eq:f1} yields components for $f_y$ and $f_z$ as follows:
\begin{eqnarray}
\label{eq:fyfzv}
\partial_t \fv_y + \frac{q}{mc}(\Bv_0 \times \fv_z + \Bv_z \times \fv_{x(0)}) +\nu \fv_y + \frac{2}{5}c \partial_x \fv_{xy} = 0\\\nonumber
\partial_t \fv_z + \frac{q}{mc}(\Bv_0 \times \fv_y + \Bv_y \times \fv_{x(0)}) +\nu \fv_z + \frac{2}{5}c \partial_x \fv_{xz} = 0.
\end{eqnarray}
Together with the MHD equations for the thermal background plasma this forms a closed system:
\begin{eqnarray}
\label{eq:maxwell}
\partial_t \Bv &=& \nabla \times (\uv \times \Bv)\\
\rho \partial_t  \uv&=&\frac{{\bf j}_{th} \times \Bv}{c}-\nabla P - \nabla \cdot \Pi - {\bf F}_R.
\label{eq:maxwell2}
\end{eqnarray}
In a thermal plasma, the stress tensor $\Pi$ is zero. We assume a cold plasma where the thermal pressure can be neglected, meaning that also $\nabla P=0$. The last term is a resistive force resulting from the collisions with the cosmic rays, and is equal but of opposite sign to the frictional force in the momentum equation for cosmic rays. 

The current in the plasma, minus the $\nabla \times B$ component, needs to balance the cosmic ray current minus the thermal component, such that:
\begin{eqnarray}
{\bf j}_{th}-\frac{c}{4\pi}\nabla\times \Bv = -{\bf j}_{cr}\nonumber
\end{eqnarray}
Writing the momentum equation for the plasma as its equivalent in terms of the cosmic rays gives:
\begin{eqnarray}
\rho \partial_t \uv &=&\frac{{\bf j}_{th} \times \Bv}{c}- {\bf F}_R\\\nonumber
&=& -\frac{{\bf j}_{cr} \times \Bv}{c}+\frac{1}{4\pi}(\nabla\times\Bv)\times\Bv + \nu \rho_{cr} (\uv_{cr}-\uv).
\end{eqnarray}
where $\rho_{cr}$ is the mass density of the relativistic particles, and where $\uv_{cr}-\uv$ is the drift speed of the cosmic rays relative to the plasma. To zeroth order, where $\uv_0=0$ this equals the shock velocity $\uv_s$. We ignore the second term on the right hand side in the second line further on. This Alfv\'enic term only becomes important on very short scales when the tension in the magnetic field lines damps the waves. The frictional force ${\bf F}_R$ has been set to be equal and opposite to the force induced by cosmic ray collisions on small scales, being: $\nu \rho_{cr} \uv_{cr}$, where we neglect the contribution of $\uv_1$. The drift velocity is related to $\fv_1$ by $\uv_{cr}f_0=\fv_1 c/3$, such that we can write the momentum equation as:
\begin{eqnarray}
\partial_t  \uv_1 =  \frac{n q}{3 m_{cr}}(\Bv_0 \times \fv_{1(1)}+\Bv_1 \times \fv_{1(0)}) +\frac{c}{3}\nu n\fv_{1(1)},
\end{eqnarray}
with $n$ the number fraction of cosmic ray protons to background protons, and where we use that $u_s f_0= f_x c/3$.
We can eliminate the magnetic field from this equation using Eq.~\ref{eq:fyfzv}:
\begin{eqnarray}
\partial_t  \uv_1&=&-\frac{nc}{3}\left(\partial_t \fv_1 + \frac{2}{5}c\partial_x \fv_{2(1)}\right).
\end{eqnarray}
Feeding this into the induction equation, we can express the perturbed magnetic field in terms of $\fv_1$ and $\fv_2$:
\begin{eqnarray}
\partial_t^2 B_y &=& -\frac{nc}{3}B_0\partial_x (\partial_t f_{y}+\frac{2}{5}v\partial_x f_{xy})\\\nonumber
\partial_t^2 B_z &=& -\frac{nc}{3}B_0\partial_x (\partial_t f_{z}+\frac{2}{5}v\partial_x f_{xz}).
\end{eqnarray}

We assume plane wave solutions of the form $\xi_\perp=\xi_1 {\rm e}^{i(kx -\omega t)}$ to derive the dispersion relation
which gives for the perturbations in the magnetic field:
\begin{eqnarray}
\label{eq:B1}
B_y = -\frac{nkcB_0}{3\omega^2}\left(\omega f_{y} -\frac{2 kc}{5} f_{xy}\right)\\\nonumber
B_z = -\frac{nkcB_0}{3\omega^2}\left(\omega f_{z} -\frac{2 kc}{5} f_{xz}\right).
\end{eqnarray}
The linearised momentum transport equation then is:
\begin{eqnarray}
\label{eq:f2linear}
(\nu -i \omega) f_y - \frac{q B_0}{mc}f_z + \frac{q f_{x(0)}}{mc} B_z  + \frac{2 ikc}{5}f_{xy} = 0\\\nonumber
(\nu-i\omega) f_z + \frac{q B_0}{mc} f_y - \frac{q f_{x(0)}}{mc} B_y + \frac{2 ikc}{5}f_{xz} = 0.
\end{eqnarray}

The relevant $f_{xy}$ and $f_{xz}$ components are derived from the next moment of the transport equation, being the heat transport or pressure tensor equation: 
\begin{eqnarray}
\label{eq:f2}
&&\frac{8\pi}{15}\left[\partial_t \{\fv_2\} + c\left(\nabla\fv_1-\frac{1}{3}\nabla \cdot \fv_1\{ {\bf I_2}\}\right)\right. \\\nonumber
&&\quad+\left. \frac{2q}{mc}\left(\Bv \times \{\fv_2\}\right) +3 \nu \{\fv_2\} + \frac{3}{7}c\nabla\cdot \{\fv_3\}\right]_2  = 0,
\end{eqnarray}
where we eliminated the $\partial_t \fv_0$ component using Eq.~\ref{eq:f0}. We use the notation as in \citet{1960Johnston}, where $\{\,\}$ is used to distinguish tensors from vectors when both appear in an equation, ${\{\bf I_l\}}$ is the $l$-th order unity tensor, and the subscript in $[ \ldots ]_l$ indicates that the equation is a summation of permutations for {\bf ijk} for each element $l!$ ways, divided by $l!$. This gives a symmetric tensor so that e.g.~$f_{xy}=f_{yx}$ with components that look like:
\begin{eqnarray}
&&\partial_t f_{xy}+\frac{c}{2}(\partial_x f_y+\partial_y f_x)\\\nonumber
&&\quad+\frac{q}{mc}(B^z f_{xx}-B^x f_{xz}+B^y f_{yz}-B^z f_{yy}) + 3\nu f_{xy}=0.
\end{eqnarray}

For now we ignore contributions to the distribution function that are third and higher order in anisotropy (we come back to the higher order equations in Sect.~\ref{sec:f3}), and set those (which includes $\partial_t \fv_2$) to zero. The non-zero contributions from Eq.~\ref{eq:f2} then are limited to:
\begin{eqnarray}
&&\frac{c}{2}\partial_x f_y - \frac{qB_0}{mc}f_{xz} + 3\nu f_{xy} = 0\\\nonumber
&&\frac{c}{2}\partial_x f_z + \frac{qB_0}{mc}f_{xy} + 3\nu f_{xz} = 0\\\nonumber
\end{eqnarray}
such that we can substitute the $\fv_2$ components in the momentum transport equation and in the perturbed magnetic field components using: 
\begin{eqnarray}
\label{eq:fxyfxz}
f_{xy} = -\frac{ik c (3 \nu f_y+\omega_g f_z)}{2((3 \nu)^2 + \omega_g^2)}\\\nonumber
f_{xz} = -\frac{i k c (3 \nu f_z-\omega_g f_y )}{2((3 \nu)^2 + \omega_g^2)},
\end{eqnarray}
where we wrote $q B_0/mc=\omega_g$ for the gyrofrequency. These components represent the transport of the perpendicular cosmic ray current.

We now have all the components to write Eq.~\ref{eq:f2linear} as a function of just $f_y$ and $f_z$ as perturbed variables: 

\begin{eqnarray}
&&\left[\nu - i\omega +\frac{k^2 c^2 3 \nu}{5((3 \nu)^2+\omega_g^2)}-\frac{i\Omega^2}{\omega^2}\left(
\frac{k^2 c^2 \omega_g}{5((3\nu)^2+\omega_g^2)}\right)\right]f_y
\\\nonumber
&&
=\left[\omega_g-\frac{k^2c^2\omega_g}{5((3\nu)^2+\omega_g^2)}+\frac{i\Omega^2}{\omega^2}\left(
i\omega-\frac{k^2c^2 3\nu}{5((3\nu)^2+\omega_g^2)}\right)\right]f_z\;,\\\nonumber\\\nonumber
&&\left[\nu - i\omega +\frac{k^2 c^2 3 \nu}{5((3 \nu)^2+\omega_g^2)}-\frac{i\Omega^2}{\omega^2}\left(
\frac{k^2 c^2 \omega_g}{5((3\nu)^2+\omega_g^2)}\right)\right]f_z
\\\nonumber
&&
= -\left[\omega_g-\frac{k^2c^2\omega_g}{5((3\nu)^2+\omega_g^2)}+\frac{i\Omega^2}{\omega^2}\left(
i\omega-\frac{k^2c^2 3\nu}{5((3\nu)^2+\omega_g^2)}\right)\right]f_y\;,
\end{eqnarray}
where $\Omega=\sqrt{k j_0 B_0/(\rho c)}$. 
\pagebreak
These can be readily combined to find the dispersion relation:
\begin{eqnarray}
\label{eq:dispf2}
\omega^2=&&
\mp\Omega^2\left(i\omega-\frac{k^2 c^2}{5(3 \nu\mp i\omega_g)}\right)
\\\nonumber &&
\left/
\left(\nu - i\omega \mp i\omega_g+\frac{k^2 c^2}{5(3 \nu\mp i\omega_g)}\right)\right..
\end{eqnarray}
The upper signs correspond to the left-handed polarisation ($f_z=if_y$) and the lower signs to the right-hand polarisation. 

\section{Evaluation of dispersion relation}
\label{sec:k0limit}

Because the growth time of the instability will always be larger than the cosmic ray crossing time ($1/kc$), the second term on the right hand side of Eq.~\ref{eq:dispf2} always dominates over the first one, and the dispersion relation can be simplified to:
\begin{eqnarray}
\label{eq:dispf2wsmall}
\omega^2=
\pm\Omega^2\left(\frac{k^2 c^2}{5(3 \nu\mp i\omega_g)}\right)
\left/
\left(\nu \mp i\omega_g+\frac{k^2 c^2}{5(3 \nu\mp i\omega_g)}\right)\right.,
\end{eqnarray}

In the short wavelength limit ($kc \gg \omega_g,\nu$), where the last term in the denominator becomes the dominant one, this method naturally retrieves the growth rate $\Omega$ that corresponds to the small-scale non-resonant instability \citep{2004Bell}. 

In the long-wavelength limit, i.e. for $k r_g \ll 1$, the dispersion relation can be reduced to: 
\begin{eqnarray}
\label{eq:dispf2ksmall}
\omega^2=&&
\pm\Omega^2\left(\frac{k^2 c^2}{5(3 \nu^2 \mp i4\omega_g\nu -\omega_g^2)}\right).
\end{eqnarray}
From this it can be readily seen that below resonance the dependency of growth rate on $k$ is very steep: $\gamma\propto k^{3/2}$. This holds for both polarisations. For which one the growth rate is more rapid depends on the relative scattering efficiency. 

The dependency that we find here is steeper than the one derived in \citet{2004Bell} in the long-wavelength limit, which was $\propto k$. The reason for the faster growth found in that paper is the integration of the distribution function over momentum, with the upper limit of the momentum set to $\infty$. This means that even for long wavelengths there was always some contribution to the instability from cosmic rays that are in resonance with the magnetic instability. In reality the maximum energy is finite and therefore on the long scales there are no resonant particles to contribute to the instability.

\begin{figure}
\begin{center}
\includegraphics[trim=60 30 40 10,clip=true,width=0.5\textwidth]{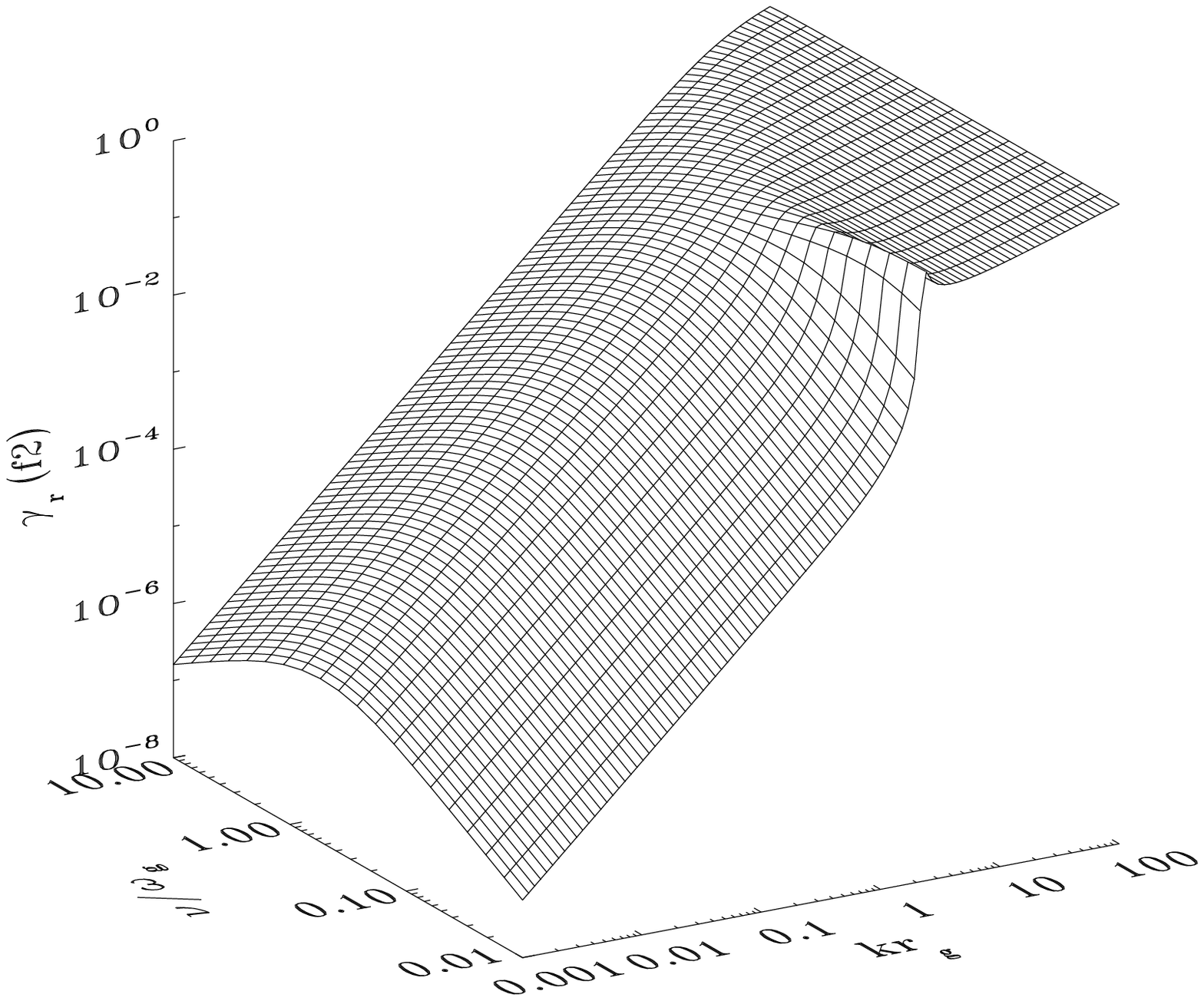}
\includegraphics[trim=60 30 40 10,clip=true,width=0.5\textwidth]{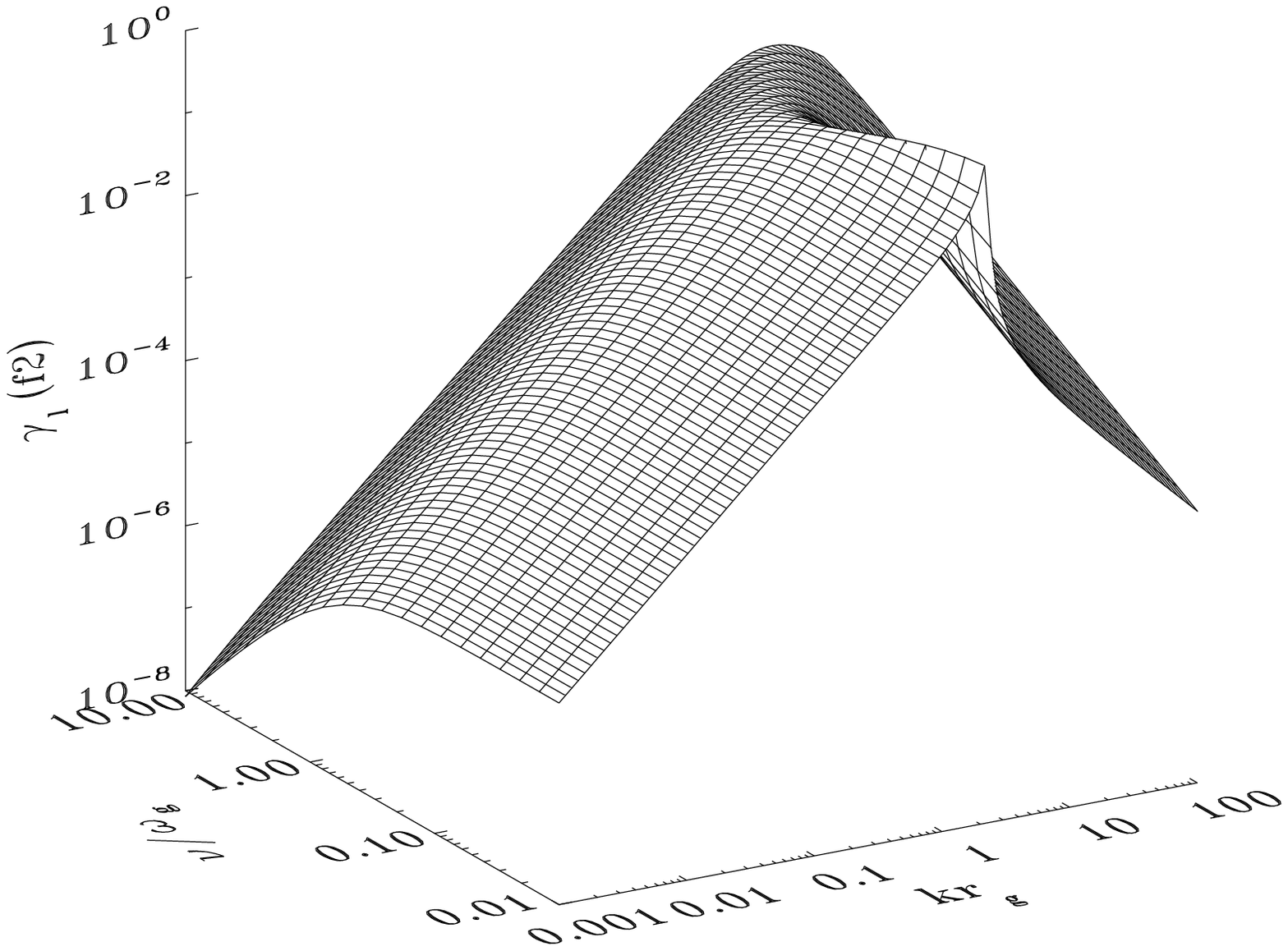}
\caption{Growth rate $\gamma$ as a function of scattering efficiency $\nu/\omega_g$ and wavenumber $k$. The upper (lower) panel represents the right (left) handed polarisation. The growth rate is normalised in units of $v_s^2/(r_g c)$.}
\label{fig:fyfzf2}
\end{center}
\end{figure}

In Fig.~\ref{fig:fyfzf2}  we plot the growth rate as a function of both the scattering efficiency and the wavenumber.  The upper panel shows the right-hand polarisation (lower signs in the equation) and the lower panel the left-hand one (upper signs). For wavenumbers below resonance the growth rate follows a $k^{3/2}$ dependency for both polarisations. Beyond resonance (on smaller scales), the growth rate converges to the small-scale non-resonant growth rate $\Omega$ for the right-handed polarisation ($\gamma \propto k^{1/2}$), whereas a steep decline for the left-handed polarisation is observed. 

\begin{figure}
\begin{center}
\includegraphics[trim=25 10 0 10,clip=true,width=0.5\textwidth]{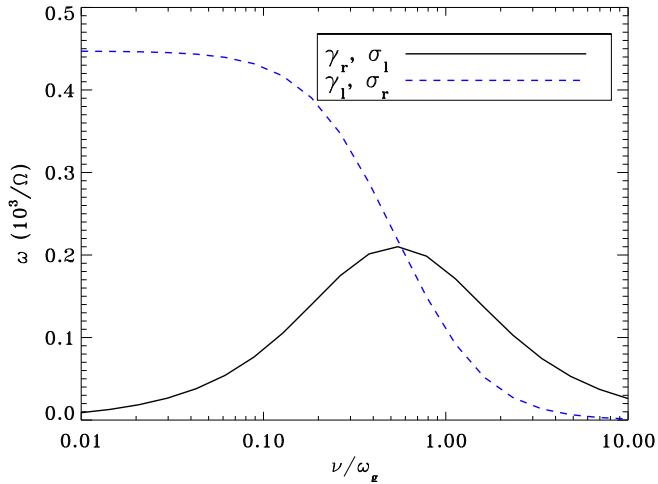}
\caption{Growth rate $\gamma$ as a function of scattering efficiency for the longest wavelength modes (normalized to $\Omega$). The solid (dashed) line represents the right (left) handed polarisation. The real part of the frequency $\sigma$ equals the growth rate of the opposite polarisation in the regime where $kr_g \leq1$.}
\label{fig:wk0_nu}
\end{center}
\end{figure}

The relative strength of the growth rate for the two polarisations on long scales depends on the relative scattering efficiency. Efficient scattering favours growth of the right-handed polarisation whereas low rates of $\nu/\omega_g$ favours the left-handed growth rate. This can be more easily seen in Fig.~\ref{fig:wk0_nu} where the growth rate is represented as a function of the scattering efficiency in the long wavelength limit. In this figure a value of $kr_g=0.001$ was assumed, and the growth rate is normalised with respect to $\Omega$. The scattering rate for which the right-handed polarisation is maximum, which is also the point where the left-handed polarisation drops below the right-handed one, is $\nu/\omega_g=\sqrt{1/3}$. This value remains the same for different $k$, as long as we evaluate it below the resonance frequency. 

We have plotted a range in scattering efficiencies between $\nu/\omega_g=[0.01, 10]$. Low values mean that the cosmic rays complete more than a full gyration before scattering, and the particles are essentially magnetised. The more frequent collisions become, the more effective diffusion occurs across field lines. Very often Bohm diffusion is used, for which the scattering frequency is of the order of the gyrofrequency. It is assumed that such a scattering efficiency is optimal in isotropising the cosmic rays and hence maximises the acceleration efficiency. Very high values of the scattering efficiency means that the particles essentially do not `see' a coherent magnetic field. This situation might arise when turbulence is very strong and disordered on the relevant length scales.

\begin{figure}
\begin{center}
\includegraphics[trim=25 10 0 10,clip=true,width=0.5\textwidth]{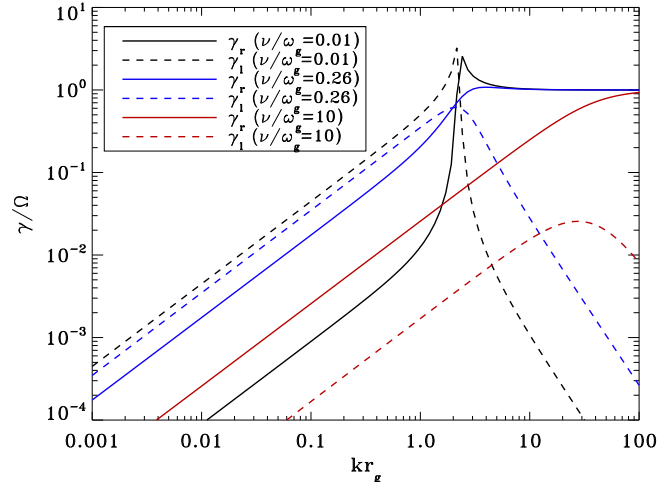}
\caption{Growth rate (normalised to $\Omega$) as a function of wavenumber for various values of the scattering rate. The solid (dashed) lines represent the right (left) handed polarisation. }
\label{fig:f2qunuvar}
\end{center}
\end{figure}

In Fig.~\ref{fig:f2qunuvar} we plotted the growth rate, normalised to $\Omega$ as a function of wavenumber, for different values of $\nu/\omega_g$. From this figure it can be seen that the growth rate of the right-hand mode at large $k$ is equal to $\Omega$, and it depends on $\nu/\omega_g$ how rapidly after resonance this convergence is reached. For low scattering efficiency the left-hand polarisation dominates, and a strong peak arises around resonance that connects the left-hand mode to the right-hand one. Evaluation of $\fv$ up to higher orders than $\fv_2$ seems to reduce this peak and eventually completely smooth it out, as will be discussed in Sect.~\ref{sec:f3}.

\section{Dissecting the mediators of the instability}
\label{sec:diff_f1f2}
In order to separate the influences of $\fv_2$ and the collision term on the instability, we look at how the instability evolves when small-scale collisions are absent or when the pressure tensor is omitted. 

\begin{figure}
\begin{center}
\includegraphics[trim=25 10 0 10,clip=true,width=0.5\textwidth]{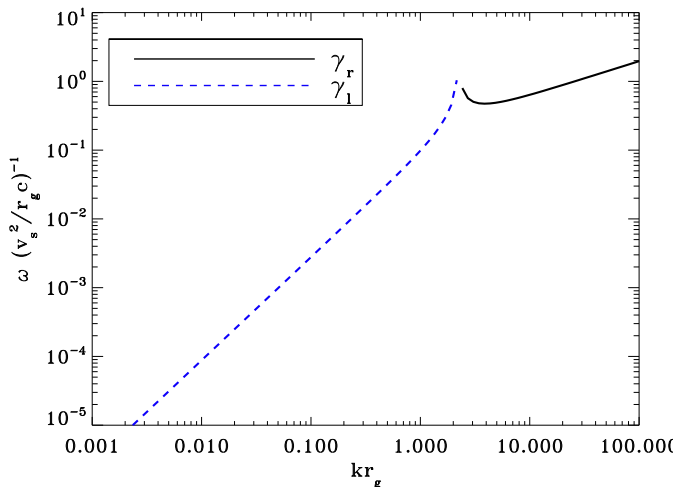}
\caption{Growth rate as a function of wavenumber when collisions are absent ($\nu=0$). The solid (dashed) line represents the right (left) handed polarisation. The right-hand polarisation is stable on scales larger than the resonant wavelength ($kr_g < 1$) whereas the left-hand polarisation is stable on small scales ($kr_g>1$).}
\label{fig:nonufyfzf2}
\end{center}
\end{figure}

The effect of neglecting collisions on short scales can be easily seen when setting $\nu/\omega_g=0$ in the dispersion relation. We plot this in Fig.~\ref{fig:nonufyfzf2}, from which we can see that both the short-scale and the long-scale instabilities are still there. However, the coupling between the right-hand and left-hand mode is completely absent. In the long-wavelength limit, the right-hand mode is stable and only the left-hand mode arises, and it retains the same $k^{3/2}$ dependency. The turning point is at resonance, after which the short-scale instability takes over, which is the same one as the 2004 Bell instability with a growth rate $\gamma \propto k^{1/2}$.

Omission of the pressure tensor, while keeping the collision term, also results in an instability, but with a much smaller growth rate. This can be seen by 
evaluating Eq.~\ref{eq:dispf2} to lowest order in $k$ (where it only appears in $\Omega=\sqrt{k j_0 B_0/(\rho c)}$), while disregarding the $k^2$ terms that originate from the inclusion of $\fv_2$. In the limit $\omega \ll \omega_g$, the dispersion relation reduces to:
\begin{eqnarray}
\omega^2=\mp\frac{i \Omega^2 \omega(\nu \pm i \omega_g)}{\nu^2+\omega_g^2}.
\end{eqnarray}

\begin{figure}
\begin{center}
\includegraphics[trim=60 30 40 10,clip=true,width=0.5\textwidth]{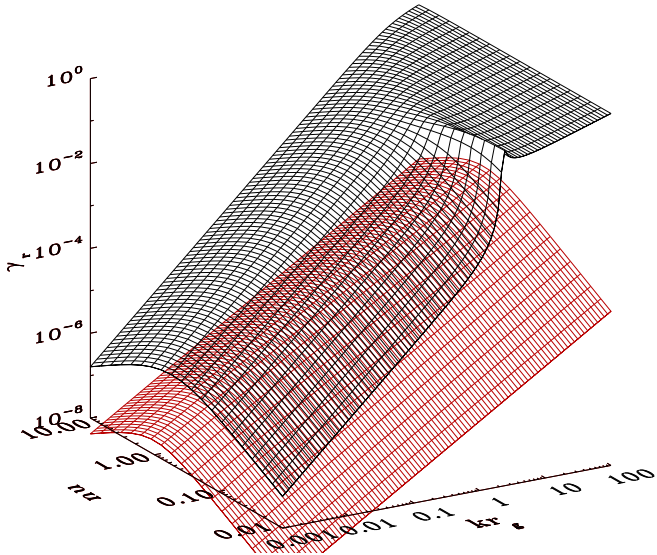}
\caption{Growth rate of the right-hand polarisation as a function of scattering efficiency and wavenumber (normalised in units of $v_s^2/(r_g c)$). The black (upper) surface shows the growth rate as it was derived earlier including the stress tensor. The red (lower) surface shows the growth rate in case the stress tensor has been ignored. The dependency on wavenumber is less steep ($\gamma \propto k$) but the growth rate remains much smaller nonetheless. Only the right-hand mode is unstable, so we have not plotted the left-hand mode. Neglect of the stress tensor results in missing vital parts of the current-driven instability. }
\label{fig:fyfzf1f2}
\end{center}
\end{figure}

When $\nu \gg \omega_g$ this would yield purely growing modes, where the growth rate (damping rate) of the right (left) hand polarisation equals:
\begin{eqnarray}
\gamma \approx \pm \frac{\Omega^2}{\nu}.
\end{eqnarray}
Only the right-hand mode is unstable. Although the dependency of the growth rate on $k$ is less steep ($\gamma \propto k$), its value is much lower compared to when $\fv_2$ was included. This can also be seen in Fig.~\ref{fig:fyfzf1f2}, in which the growth rate with and without including $\fv_2$ is compared. 

In the limit where $\nu \ll \omega_g$ both modes are stable. In the absence of collisions (in addition to omission of the pressure tensor) the dispersion relation reduces to:
\begin{eqnarray}
\omega=\frac{\Omega^2}{\omega_g},
\end{eqnarray}
and the system is stable.

This illustrates that the inclusion of the stress tensor (or momentum transport) forms an integral part of the long wavelength instability. The strong dependency of $\omega$ on $k$ in the dispersion relation (Eq.~\ref{eq:dispf2}) makes the $k^2$ term dominant. This term arised from the addition of $\fv_2$,  
the pressure tensor. 
Omission of this term will result in an {\em unphysical} result that neglects an important mediator of the instability. 

\section{Growth of the instability around resonance}
\label{sec:resonance}

In the previous section we have seen that the diffusion term is needed to mediate the instability between the two polarisations. On scales shorter than the resonance wavelength, the dominant mode is the right-hand polarisation. Growth in this regime is rapid, and it has been shown to amplify the magnetic field to many times the background field in the nonlinear stage. The filamentary structures that arise in the nonlinear regime grow as a result of the $j \times B$ force. Whether the largest growth around resonance arises due to the nonlinear development of the rapid small-scale instability, or due to the linear growth directly resulting from the resonant streaming instability, remains uncertain. Small-scale structures evolve to larger scale structures as was shown in \citet{2005Bell}.

For scales larger than the resonance wavelength we have demonstrated that the left-hand polarisation dominates the instability as long as $\nu/\omega_g < \sqrt{1/3}$. Due to the longer coherence time of the $\fv_{1(1)} \times \Bv_1$ force, the scattering of positively charged cosmic rays is more efficient for left-hand polarised magnetic fluctuations \citep{1974Wentzel}. Therefore, even though the growth rate is smaller, the resulting confinement of cosmic rays may still be quite efficient for longer wavelengths, in effect for cosmic rays with higher energies than the dominant ones that drive the instability. 

The local peak in the growth rate at the resonant wavelength reinforces the importance of the resonant instability discussed for example by \citet{1967Lerche, 1969KulsrudPearce, 1974Wentzel, 1975Skillingc}. Simulations that can treat scales both shorter and longer than resonance will be needed to determine the relative contribution of the various instabilities.

\section{Including higher orders of the distribution function}
\label{sec:f3}
In the derivation for the dispersion relation we evaluated the anisotropy of the distribution up to $\fv_2$. In Section~\ref{sec:diff_f1f2} we illustrated the difference in the dispersion relation when the pressure tensor anisotropy is neglected. Because of the major differences, it will be beneficial to determine what happens to the instability when we include anisotropy terms of higher order moments. 

In this section we evaluate the dispersion relation for the next order in $\fv$ ($\fv_3$) and show that no major differences arise in comparison to neglecting it. The higher order transport equations are derived as described in Sect.~\ref{sec:method}. The third moment yields the pressure transport or heat tensor equation:
\begin{eqnarray}
&&\frac{8\pi}{35}\left[\partial_t\{\fv_3\}+c(\nabla\{\fv_2\}-\frac{2}{5}\nabla\cdot\{\fv_2\}\{{\bf I_3}\})
\right.\\\nonumber
&&\quad+\left.\frac{3q}{mc}(\Bv\times \{\fv_3\})+6 \nu\{\fv_3\} +\frac{4}{9}c\nabla\cdot\{\fv_4\}\right]_3=0.
\end{eqnarray}

In the pressure tensor equation (Eq.~\ref{eq:f2}) we now have to take into account the contributions from $\fv_3$ and the time derivative in $\fv_2$, being of the same order as $\fv_3$. The contributions from the heat transport equation then are:
\begin{eqnarray}
&&(3\nu-i\omega) f_{xy}-\omega_g f_{xz}+\frac{3ikc}{7}f_{xxy} + \frac{ikc}{2} f_y  = 0\\\nonumber
&&(3\nu-i\omega) f_{xz}+\omega_g f_{xy}+\frac{3ikc}{7}f_{xxz} + \frac{ikc}{2} f_z  = 0.
\end{eqnarray}

As can be seen from evaluating $f_{xy}$ and $f_{xz}$, where now we do include components of $\fv_3$, we only need to determine $f_{xxy}$ and $f_{xxz}$ in order to get the dispersion relation. 
These can be derived to be:
\begin{eqnarray}
\label{eq:dtf3}
\partial_t f_{xxy}+\frac{8}{15}c\partial_x f_{xy} - \omega_g f_{xxz} + 6 \nu f_{xxy}=0\\\nonumber
\partial_t f_{xxz}+\frac{8}{15}c\partial_x f_{xz} + \omega_g f_{xxz} + 6 \nu f_{xxz}=0,
\end{eqnarray}
where we neglect terms that are of higher order than $\fv_3$, being its $\partial_t$ component and  $\fv_4$.
Eqs.~\ref{eq:dtf3} are used to solve for $f_{xy}$ and $f_{xz}$, which can be substituted in the equation for $\fv_1$ to give the modified dispersion relation:
\begin{eqnarray}
\label{eq:dispf3}
&&\omega^2=\\\nonumber
&&\qquad\mp\Omega^2\left(i\omega-\frac{k^2 c^2}{
5\left(3\nu-i\omega\mp i\omega_g+
\frac{8k^2c^2}{
35(6\nu\mp i\omega_g)
}\right)
}\right)\\\nonumber&&
\left /
\left(\nu - i\omega \mp \omega_g+\frac{k^2 c^2}{
5\left(3\nu-i\omega\mp i\omega_g+
\frac{8k^2c^2}{
35(6\nu\mp i\omega_g)
}\right)
}\right)\right..
\end{eqnarray}

\begin{figure}
\begin{center}
\includegraphics[trim=60 30 40 10,clip=true,width=0.5\textwidth]{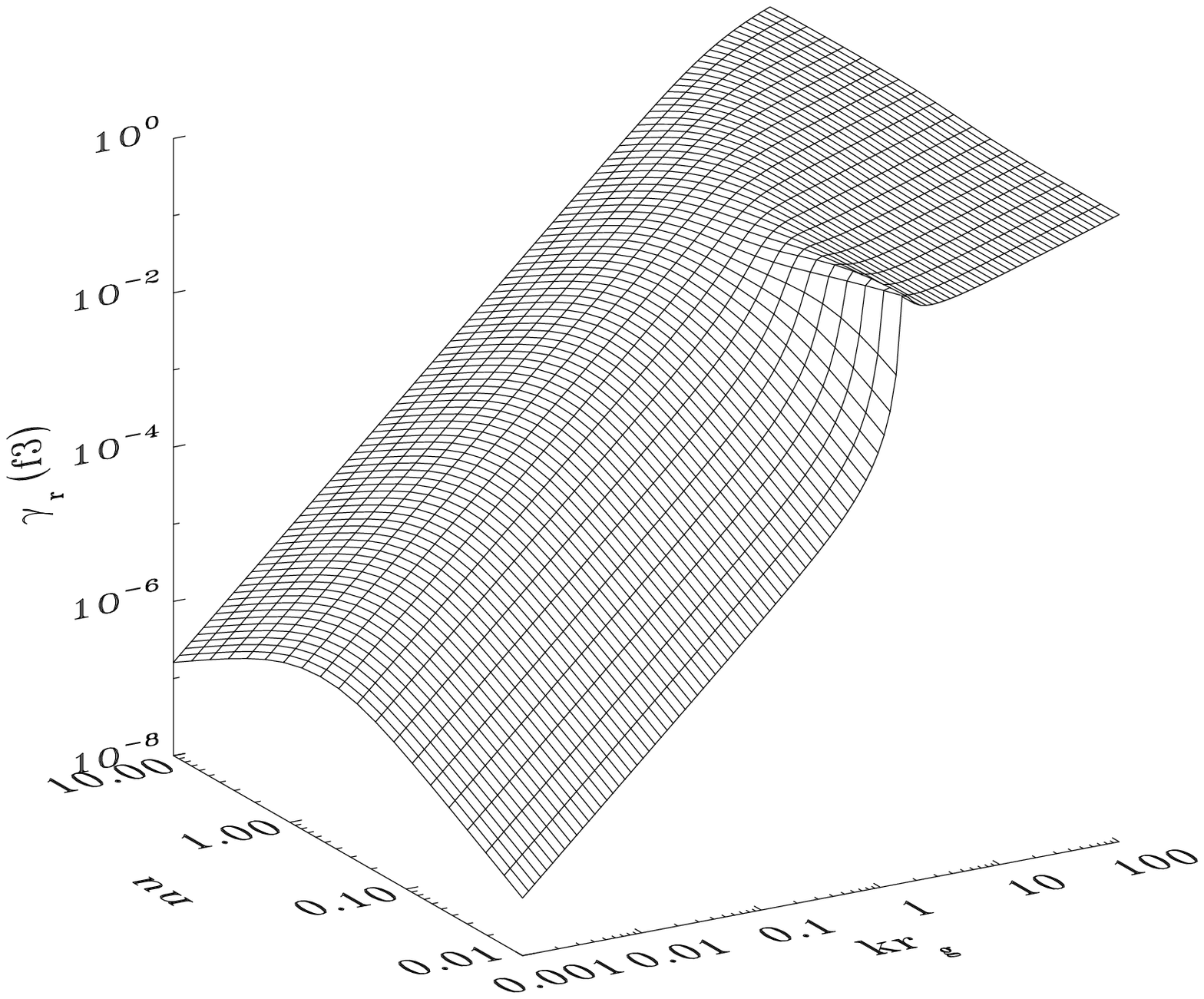}
\includegraphics[trim=60 30 40 10,clip=true,width=0.5\textwidth]{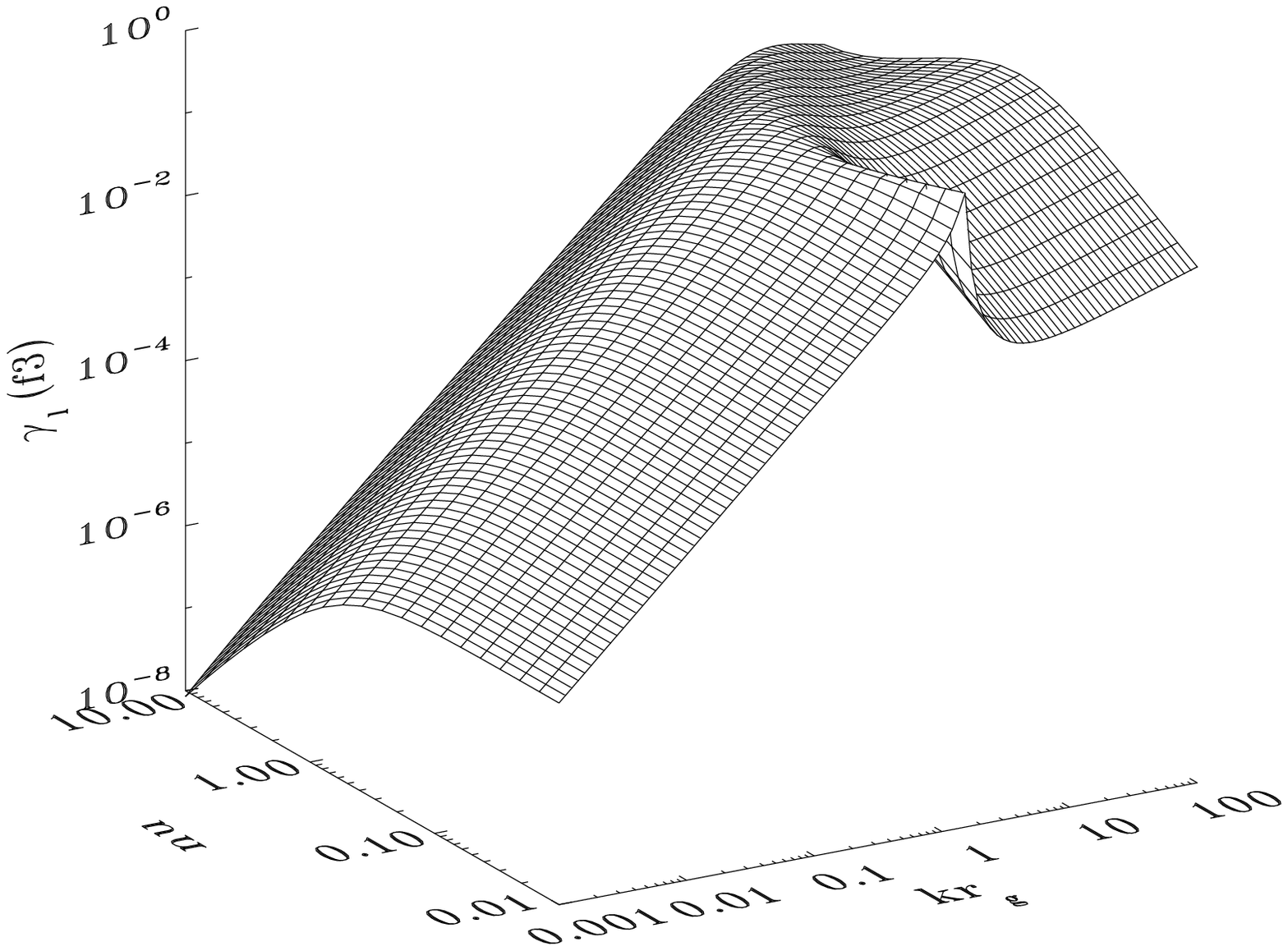}
\caption{Growth rate for the dispersion relation up to orders of $\fv_3$ as a function of scattering efficiency and wavenumber. The upper (lower) panel represents the right (left) handed polarisation. The growth rate is normalised in units of $v_s^2/(r_g c)$.}
\label{fig:wf1f3}
\end{center}
\end{figure}

Fig.~\ref{fig:wf1f3} shows the growth rate for the dispersion relation up to $\fv_3$ as a function of both $\nu/\omega_g$ and $k$. On the long scales there is no noticeable difference between this result and the one ignoring $\fv_3$. The peak around resonance softens a bit and moves to slightly shorter wave numbers, closer to $kr_g=1$. We have evaluated the dominant terms for subsequent orders in $\fv$ (up to $\fv_{10}$) and found that this trend seems to continue when higher orders of $\fv$ are included. If the trend continues as we expect, the transition around resonance smoothens out completely. 

A curious effect arises for the left-hand mode in the short-wavelength limit. Rather than steeply declining with wave number, it increases when the highest order of $\fv$ that is included is odd (i.e.~$\fv_3$ or $\fv_5$). When the highest order of $\fv$ that is included is an even number, the growth rate asymptotically decreases with wave number. From the dispersion relation it can be seen how this effect arises: When the highest order is an even number, the orders of $k$ in the nominator are exactly matched in the denominator and thus cancel, whereas this is not the case when the highest order is an odd number. In the analysis by \citet{1983Achterberg} the growth rate in this regime is shown to decline for high $k$, which leads us to believe that the real trend is captured when the highest order is an even order of $\fv$ ($\fv_2$, $\fv_4$), which has a decreasing growth rate with wave number for the left-hand mode. 

\section{Discussion and conclusions}
\label{sec:discussion}
We have derived the dispersion relation for an instability that is driven by the zeroth order cosmic ray current. The long-wavelength component is mediated by the stress tensor, and coupling between the left- and right-hand modes occurs when scattering of cosmic rays off small scale turbulence is taken into account. 

The streaming of cosmic rays ahead of the shock induces a return current in the plasma that triggers instabilities from large scales down to the Alfv\'enic regime on very small scales at which the magnetic tension becomes efficient in damping the oscillations. We did not include the Alfv\'en term (magnetic tension), which only becomes of importance beyond $k r_g > 10^3$ for our parameters. 

On small scales our analysis reproduces the hydrodynamical non-resonant instability as described in \citet{2004Bell}.  
The growth rate at small scales is very high and the amplified magnetic field can reach many times its zeroth order strength. In an environment such as that upstream of a SNR blast wave, the efficient generation of magnetic field amplification on small scales is expected to increase the scattering efficiency of the lower energy cosmic rays. We find that this in turn seems to couple the left- and right-hand modes, resulting in a small left-handedly polarised contribution to the short-scale instability, in addition to the dominant right-handed mode.  

On long scales, predominantly the left-hand mode is unstable. This effect only arises when anisotropy of order $\fv_2$ (and higher) is taken into account. The long-wavelength mode behaviour already converges when orders of $\fv_2$ are included, as we find that higher-order anisotropy terms do not alter the instability significantly. Inclusion of the collision term
results in coupling between the two polarisations such that both modes are also unstable in the long-wavelength limit. 
The growth rate of the long-wavelength part of the instability is a function of $k^{3/2}$, which results in a rapid decline of the growth rate with wavelength. In a recent paper, \citet{2011Bykovetal} found a larger growth rate in this regime using a different method that includes the ponderomotive force resulting from the instability on short scales.

Since we are assuming a mono-energetic beam of particles not all the effects around the resonance points are captured in this approach. However, since the number of cosmic rays steeply declines with energy, it can be expected that the instability is dominated by the cosmic rays with the lowest energy in the local plasma, being a function of distance ahead of the shock. 
Including higher orders of $\fv$ shows converging behaviour in the resonance region, and the peaks that arise as a result of just including up to $\fv_2$ smooth out. 

Alternatively, the magnetic fluctuations in the resonance regime could be dominated by filaments resulting from the short-scale nonresonant instability that have grown in structure to length scales that are in resonance or beyond.
It will be important to evaluate the non-linear regime and saturation levels to determine the actual efficiency of this instability in confining high energy cosmic rays. 

What happens farther upstream, in the region where only the high-energy cosmic rays diffuse out to, while the lower-energy cosmic rays are confined closer to the shock region, is uncertain. 
If effective diffusivity only operates when a significant fraction of the low-energy cosmic rays is around, the two regimes may not couple, causing the long-wavelength instability to occur exclusively in the left-handed variety and vice versa for the short-scale instability. 

An interesting result is that the dominant mode on scales larger than the gyroradius is the left-hand polarised one. This polarisation is more efficient in scattering cosmic rays since the perturbed $j \times B$ force continues to act in the same direction over longer scales. The deflection therefore is larger, and scattering more efficient. Therefore, even though the growth rate of this mode is smaller, the long-wavelength instability may still be very effective in confining cosmic rays  
that have gyroradii larger than the driving cosmic rays.

\section*{Acknowledgements}
This research was supported by the UK Science Technology and Facilities Council grant ST/H001948/1. The authors thank Bram Achterberg for pointing out an inconsistency in an early draft of this paper. We would also like to thank Brian Reville and Andrei Bykov for interesting discussions on the topic.

\bibliography{../../adssample}
\label{lastpage}
\end{document}